\documentclass[fleqn,usenatbib]{mnras}

\usepackage{newtxtext,newtxmath}
\usepackage{adjustbox}
\usepackage{ragged2e}
\usepackage{multirow}
\usepackage{graphicx}

\usepackage[T1]{fontenc}

\DeclareRobustCommand{\VAN}[3]{#2}
\let\VANthebibliography\thebibliography
\def\thebibliography{\DeclareRobustCommand{\VAN}[3]{##3}\VANthebibliography}

\usepackage{graphicx}	
\usepackage{amsmath}	

\title[Study of central intensity ratio]{Study of central light distribution in nearby early-type galaxies hosting nuclear star clusters}

\author[K. Sruthi $\&$ C. D. Ravikumar]{
K. Sruthi,$^{1}$\thanks{E-mail: sruthiyatheendradas@gmail.com, cdr@uoc.ac.in}
and C. D. Ravikumar$^{1}$
\\
$^{1}$Department of Physics, University of Calicut, Malappuram 673635, India\\
}

\date{Accepted XXX. Received YYY; in original form ZZZ}

\pubyear{2023}

\begin{document}
\label{firstpage}
\pagerange{\pageref{firstpage}--\pageref{lastpage}}
\maketitle

\begin{abstract}
We present analysis of 63 nearby  ($<$ 44 Mpc) early-type galaxies hosting nuclear star clusters using the recently discovered parameter Central Intensity Ratio (CIR$_I$) determined from near-infra-red (3.6 $\mu$m) observations with the Infra-red-array-camera of \emph{Spitzer} space telescope. The CIR$_I$, when combined with filters involving age and $B-K$ colour of host galaxies, helps  identify two distinct classes of galaxies hosting nuclear star clusters. This is independently verified using Gaussian Mixture Model. CIR shows a positive trend with  faint, low mass, and blue galaxies in the sample, while the opposite is true for bright, high mass, and red galaxies, albeit with large scatter.  The variation of CIR$_I$ with central velocity dispersion, absolute B band magnitude, dynamical mass, and stellar mass of host galaxies suggests that the mass of nuclear star clusters increases with that of host galaxies, for faint, low mass, young and blue galaxies  in the sample.  In bright, high-mass, old and red  galaxies, on the other hand, the evolution of nuclear star clusters  appears complex, with no apparent trends.The analysis also reveals that redder galaxies ($B-K > 3.76$) are more likely to be dominated by the central black-hole than the nuclear star clusters, while for bluer galaxies ($B-K < 3.76$) in the sample the situation is quite opposite. 
\end{abstract}




\section{Introduction}
The evolution of galaxies is closely connected with interactions and mergers \citep{2012MNRAS.425.2313T}.  The minor and major interactions \citep[e.g.][]{2007AJ....134..527W,2009ApJ...704..324R,2012A&A...539A..45L} have significant effects on the star formation, morphology, kinematics, and physical conditions of gas, both in the central and outer regions of galaxies. Large quantities of gas can be funnelled to the central region of galaxies during major mergers, triggering enormous star formation and rapid fuelling of central black holes \citep{2018MNRAS.480..947D}. Due to these reasons,  the very innermost regions of galaxies provide examples of extreme astrophysical environments, where numerous complex mechanisms operate simultaneously that have direct as well as indirect roles in the formation and evolution of galaxies. The properties of nuclear regions are thought to be linked to the formation history of galaxies where the collapse of gas and merging of galaxies can significantly influence the creation of different types of galaxies \citep{2014A&A...568A..89G}.  As a result, the very central regions of galaxies host fascinating objects like supermassive black holes (hereafter SMBHs) and active galactic nuclei (AGN) whereas nuclei of some galaxies are undergoing extreme star formation at their centres in the form of central starbursts, and extreme stellar densities \citep{2020ApJ...900...32P}. All these are likely connected to the global properties of their host galaxies, and the co-evolution of various components is invoked to account for the observed correlations \citep{2014A&A...568A..89G}. For example, masses of SMBHs have been shown to correlate with a range of host-galaxy properties \citep[e.g.][]{2013ARA&A..51..511K,2016ASSL..418..263G,2016ApJ...818...47S,2017MNRAS.471.2187D,2018MNRAS.477.2399A,2021MNRAS.500.1343S}.

Over the past two decades, high-resolution observations with the \textit{Hubble} space telescope have revealed that, in some galaxies, the central SMBH is surrounded by a massive, very compact, star cluster composed of stars up to 10$^{8}$ in number, commonly known as nuclear star cluster (NSC). These massive stellar clusters reside at the photometric and dynamic centres of most intermediate and low luminosity galaxies of all Hubble types \citep{2002AJ....123.1389B,2014MNRAS.445.2385D,2014MNRAS.441.3570G}. NSCs are known to be the densest stellar systems in the universe with millions of solar masses packed within the central few parsecs of a galaxy and therefore host unique stellar dynamics \citep{2016ASSL..418..107C}. These objects are identified as a clear over-density in stars above the inwardly extrapolated light profile of a galaxy within the central 50 pc \citep{2020A&ARv..28....4N}. NSCs are sited at the bottom of the potential well \citep{2014MNRAS.444.3738A} and contain old as well as young stellar  populations \citep[e.g.][]{2006AJ....132.1074R,2012AdAst2012E..15N}.  NSCs and SMBHs are proposed to be related to their host galaxies in similar ways, which activated a new wave of interest in the study of NSCs and their role in the growth of SMBH \citep[e.g.][]{2006AJ....132.1074R,2006ApJ...644L..17W,2006ApJ...644L..21F}. Galaxies and NSCs display a variety of scaling relations suggesting that, their formation is intricately linked to that of their host galaxy \citep[e.g.][]{2014MNRAS.441.3570G,2016MNRAS.457.2122G}. However, unlike SMBHs, the formation history of NSCs is directly visible through their stellar populations which allows for detailed studies of the mass accretion history that takes place in galactic nuclei \citep{2018MNRAS.480.1973K}. Though NSCs are observed in galaxies with every type of the Hubble sequence, their modes of formation and evolution are still under debate \citep{2014MNRAS.444.3738A}. Two different mechanisms are proposed for the formation of NSCs, infall of star clusters that formed elsewhere in the host galaxy \citep[e.g.][]{1975ApJ...196..407T,2008ApJ...681.1136C,2013ApJ...763...62A} and \emph{in situ} formation and build-up through star formation following the accretion of gas in the centre of galaxies \citep{2007PASA...24...77B}. There is a need to accurately determine scaling relations involving NSCs to improve our understanding of the possible evolution of NSCs with their host-galaxies \citep{2020ApJ...900...32P}. Studying how NSCs form and how they are related to the growth of central massive black holes and their host galaxies is crucial for our understanding of the evolution of galaxies and the physics that have shaped their central components \citep{2015ApJ...806L...8A}.

NSCs and SMBHs are sometimes jointly referred to as central massive objects \citep[CMO;][]{2006ApJ...644L..21F, 2014MNRAS.444.3738A} whose presence is connected to the host galaxy mass; galaxies with mass above 10$^{10}$ M$\odot$ usually host SMBH while lighter galaxies have a well resolved nuclear star cluster. Moreover, a transition region exists for galaxies with a mass between 10$^{8}$ M$\odot$ and 10$^{10}$ M$\odot$ where both the objects co-exist. It seems that there is a continuous sequence from NSC-dominated galaxies to SMBH-dominated galaxies as the galaxies grow bigger \citep{2016MNRAS.456.2457A}. The formation of NSCs may also be connected to the formation of SMBH in the centre, and these objects could therefore trace the evolutionary history of the host galaxy formation. For instance, NSCs sit on the low mass extrapolation of the scaling relation between the SMBH mass and the total mass of the host galaxy \citep{2014A&A...568A..89G}. However, interactions with SMBHs have the potential to destroy NSCs or inhibit NSC growth in most massive galaxies \citep{2021A&A...650A.137F}.  Very few NSCs are observed in galaxies harbouring very massive SMBHs at their centres; this could be an indication of a physical connection between the presence (or absence) of NSCs and SMBHs in galaxies \citep{2016MNRAS.456.2457A}. The relationship between NSCs and SMBHs is not well understood as the size of a sample of galaxies with both NSC and a central BH is quite limited \citep{2018ApJ...858..118N}. Galaxies with masses lower than 10$^{10}$ M$_{\odot}$ show clear evidence for nucleation, but little evidence for an SMBH. Contrarily, galaxies with masses above $\sim$ 10$^{11}$ M$_{\odot}$ are dominated by SMBH but show no evidence for nucleation \citep{2015ApJ...806L...8A}. Therefore, it is an open question whether NSCs are an essential ingredient in (or an intermediate step towards) the formation of a supermassive black hole in the galaxy nucleus \citep{2014MNRAS.441.3570G}.

The co-evolution of nuclear star clusters and central black holes is presently a very active field of research. In this scenario, we explore the connection between NSCs, SMBHs and host galaxy properties with the aid of central intensity ratio \citep[CIR;][]{2018MNRAS.477.2399A,2020RAA....20...15A,2021MNRAS.500.1343S}. The central intensity ratio (defined in section \ref{sec:central}) is a photometric parameter that measures the variation of light intensity at the very centre of the (projected) galaxy image. Simple Monte Carlo simulations involving CIR show remarkable stability against variations in distance and orientation of ellipsoidal systems in the nearby universe \citep{2018MNRAS.477.2399A,2020RAA....20...15A}.  The straightforward photometric definition of CIR makes it sensitive to any addition (or subtraction) to the light near the centre of a galaxy.  The addition of light is possible in systems with high star formation near the centre while quenching via feedback from the central black hole can reduce the CIR value \citep{2021MNRAS.500.1343S}. The various correlations shown by CIR in field early-type galaxies (hereafter ETGs) \citep{2021MNRAS.500.1343S} explored the co-evolution scenario in galaxies. The CIR is also found to contain information about the star formation near the central region of galaxies \citep{2020RAA....20...15A}. Therefore, the CIR can be used as an ideal tool in studying galaxies hosting NSCs.

This paper is organized as follows. Section \ref{sec:data} describes the sample selection, observations, data reduction, and calculation of CIR$_{I}$. Section \ref{sec:results} presents variations of CIR$_{I}$ with different host galaxy properties. We present a discussion of results and conclusion in Section \ref{sec:discussion}.
\section{The Data}
\label{sec:data}
We constructed a sample of 80 nearby early-type galaxies hosting nuclear star clusters based on the availability of \emph{Spitzer}/IRAC 3.6$\mu$m observations adopted from literature. We avoided 15 galaxies from the above sample due to their relatively small size which could affect the determination of CIR$_I$. Also, we avoided two galaxies NGC 205 and M32 as they are pretty close and tidally influenced by the nearby galaxy M31. Thus the sample includes 63 nearby ETGs hosting NSCs composed of 26 ellipticals and 37 lenticulars. The sample along with the references is listed in Table \ref{tab1}. 

\subsection{Central intensity ratio}
\label{sec:central} 
Central intensity ratio \citep[CIR;][]{2018MNRAS.477.2399A} is defined as,
\begin{equation}
CIR = \frac{I_{1}}{I_{2} - I_{1}} = \frac{10^{0.4(m_{2}-m_{1})}}{1-10^{0.4(m_{2}-m_{1})}}
\end{equation} 
where \emph{I$_{1}$} and \emph{I$_{2}$} are the intensities and \emph{m$_{1}$} and \emph{m$_{2}$} are the corresponding magnitudes within the inner and outer apertures respectively. Aperture photometry (MAG$\_$APER) using Source extractor \citep[SEXTRACTOR,][]{1996A&AS..117..393B} was carried out at the photometric centre of the galaxy image. CIR measured in the optical band (hereafter CIR$_V$) used the inner and outer apertures of 1.5 and 3 arcsec respectively. For the present study, we have chosen the inner and outer radii as 12 and 24 arcsec, respectively for the calculation of CIR$_{I}$. The inner radius is selected to contain the effects of the point spread function (PSF) in the \textit{Spitzer} images and the outer radius is chosen to be smaller than the half-light radii of the sample galaxies. This ensures that the estimation of CIR is reasonably stable against differences in size and surface brightness profile \citep{2018MNRAS.477.2399A} of and distance to galaxies. However, a direct comparison of CIR$_V$ and CIR$_I$ is not possible as there are large differences in the PSF, depth and aperture sizes used. 

The simple definition of CIR helps avoid any dependence on a form following central intensity, \textit{I}(0) (i.e. surface brightness at a radial distance r, \textit{I}(r) = \textit{I}(0) \textit{f}(r) where \textit{f}(r) is a function of r). On the other hand, the definition boosts any addition to (or subtraction from) the central intensity  \textit{I}(0) \citep{2018MNRAS.477.2399A,2020RAA....20...15A,2021MNRAS.500.1343S}. NSCs are very tiny objects near the centre covering around 10 parsecs only. CIR measures the total projected light along the centre and hence the smaller the apertures the better for identifying variations in flux due to NSCs.





\begin{table*}
	\centering
	\caption{Table \ref{tab1} lists the properties of the sample galaxies. Name of the galaxy with its environment adopted from NED (column 1) where `*'(star) symbol denotes galaxy from dense environment whereas `+'(plus) symbol denotes galaxy from rare environment, References from which the sample galaxies are adopted (2) where N20 - \citet{2020A&ARv..28....4N}, SG13 - \citet{2013ApJ...763...76S}, F21 - \citet{2021A&A...650A.137F} and P20 - \citet{2020ApJ...900...32P}, Morphological type (3) taken from \citet{2013ApJ...763...76S} and NED, Distance of the galaxy (4) adopted from NED, for the galaxy NGC 5389, distance is adopted from \citet{2013MNRAS.436.3135E}, for NGC 4552,the distance is taken from \citet{2005ApJ...631..809X}, CIR computed in the near-infra-red band (5), uncertainty in the measurement of CIR$_I$ (6), Mass of central black hole (7) with references where a -\citet{2020A&ARv..28....4N}, b - \citet{2019MNRAS.484..794G}, c - \citet{2013ARA&A..51..511K}, d - \citet{2011MNRAS.410.1223R}, e - \citet{2019MNRAS.490..600D} , f - \citet{2009MNRAS.399.1839K}, g - \citet{2020ApJ...888...37D}, h - \citet{2006MNRAS.370..559S}, i - \citet{2021MNRAS.500.1933S}, j - \citet{2022MNRAS.509.2920N}, k -\citet{2012ApJ...753...79W}, l - \citet{2015Ap&SS.356..347M}, m - \citet{2021ApJ...907....6Z}, Age of the galaxy (8) adopted from \citet{2009MNRAS.394.1229C}, \citet{2014ApJ...783..135A} and \citet{2016MNRAS.460.4492D}, Dynamical mass (9) of the galaxy taken from \citet{2013ApJ...763...76S} and \citet{2016MNRAS.460.4492D}, Central velocity dispersion of the galaxy (10) and absolute B band magnitude of the galaxy (11) adopted from HyperLEDA, Stellar mass of the galaxy (12) taken from references listed in column 2, $B-K$ colour of the galaxy (13) calculated using apparent total B and K magnitudes obtained from HyperLEDA, Mass of the nuclear star cluster (14) of the galaxy adopted from references listed in column 2.}
	  
	\label{tab1}
	\begin{tabular}{lccccccccccccr}
	\hline
Galaxy   & Ref. & Type & Dist. & CIR$_I$ & $\bigtriangleup_{CIR_I}$  & M$_{BH}$  & Age & M$_{dyn.}$ & $\sigma$  & M$_B$     & M$_{str}$ & $B-K$ & M$_{NSC}$\\
         &      &    & (Mpc)      &         &      & (log M$\odot$)     & (Gyr) &          (log M$\odot$) &  (kms$^{-1}$)      & (mag)       & (log M$\odot$) &     (mag) & (log M$\odot$)\\
 \hline
IC1459$^{+}$   & N20  & E    & 28.70 & 0.93 & 0.01 & 9.39$^a$ & -     & 11.67 & 296.11 & -21.40 & 11.10 & 4.12 & 7.69 \\
IC3032$^{*}$   & SG13 & dE   & 15.00 & 0.52 & 0.09 & 4.82$^b$ & 8.30  & 9.31  & 20.80  & -15.79 & 8.70 & 2.97 & 5.66\\
IC3065$^{*}$   & SG13 & S0   & 17.15 & 0.63 & 0.05 & 5.21$^b$ & 5.30  & 9.30  & -      & -16.84 & 9.30 & 3.37 & 5.70\\
IC3292$^{+}$   & SG13 & dS0  & 15.59 & 0.80 & 0.05 & 4.78$^b$ & 10.60 & 9.70  & 32.10  & -16.09 & 8.90 & 3.27 & 6.13\\
IC3381$^{*}$   & SG13 & dE   & 17.16 & 0.55 & 0.06 & 4.94$^b$ & 5.70  & 9.40  & 40.35  & -17.27 & 9.30 & 2.72& 6.73\\
IC3468$^{*}$   & SG13 & E    & 16.27 & 0.62 & 0.05 & 4.60$^b$ & 4.80  & 9.50  & 32.95  & -17.18 & 9.40 & 3.14 & 6.66\\
IC3470$^{*}$   & SG13 & dE   & 17.50 & 0.66 & 0.05 & 5.43$^b$ & 6.20  & 9.50  & 52.40  & -16.70 & 9.20 & 3.39& 6.80\\
IC3586$^{*}$   & SG13 & dS0  & 20.04 & 0.69 & 0.07 & 5.35$^b$ & 6.50  & 9.30  & 24.40  & -16.77 & 8.80 & 2.33 & 5.81\\
IC3735$^{*}$   & SG13 & dE   & 15.44 & 0.56 & 0.08 & 4.83$^b$ & 7.70  & 9.60  & 37.00  & -16.69 & 9.10  & 3.01& 6.54\\
IC3773$^{+}$   & SG13 & dS0  & 17.42 & 0.71 & 0.05 & 5.80$^b$ & 4.90  & -     & 59.49  & -17.34 & 9.30 & 3.06 & 6.10\\
IC798$^{*}$    & SG13 & E    & 18.90 & 0.44 & 0.07 & 4.76$^b$ & 9.94  & 9.03  & -      & -16.11 & 9.00 & 3.19& 6.93\\
IC809$^{*}$    & SG13 & dE   & 15.77 & 0.62 & 0.05 & 5.15$^b$ & 5.00  & 9.70  & 37.00  & -16.68 & 9.40 & 3.78 & 6.79\\
NGC0404$^{+}$  & N20  & S0   & 3.13  & 0.93 & 0.02 & 5.74$^a$ & -     & 10.04 & 34.59  & -16.43 & 8.86 & 3.66 & 7.08\\
NGC1023$^{+}$  & SG13 & SB0  & 10.84 & 0.81 & 0.01 & 7.62$^c$ & 12.30 & 11.00 & 197.10 & -20.61 & 10.80 & 3.31 & 6.64\\
NGC1332$^{+}$  & N20  & S0   & 21.90 & 0.82 & 0.01 & 9.18$^d$ & -     & -     & 294.62 & -20.74 & 11.69 & 4.13 & 7.14\\
NGC1336$^{+}$  & F21  & S0   & 17.30 & 1.21 & 0.03 & -    & -     & -     & 97.67  & -18.21 & 9.97 & 3.28& 8.74\\
NGC1375$^{+}$  & F21  & S0   & 16.20 & 0.99 & 0.03 & -    & -     & 9.76  & 64.56  & -18.39 & 9.76  & 3.54 & 8.37\\
NGC1381$^{+}$  & F21  & S0   & 17.50 & 0.96 & 0.01 & -    & -     & 9.75  & 147.14 & -19.34 & 10.35 & 3.97& 8.45\\
NGC1389$^{+}$  & F21  & S0   & 16.80 & 0.95 & 0.02 & -    & -     & -     & 124.57 & -19.25 & 10.52 & 3.77& 8.15\\
NGC1399$^{+}$  & SG13 & E    & 17.71 & 0.64 & 0.01 & 8.95$^e$ & -     & 11.70 & 332.23 & -21.22 & 11.30 & 4.10 & 6.81\\
NGC2549$^{+}$  & SG13 & SB0  & 16.78 & 1.04 & 0.01 & 7.15$^f$ & 4.80  & 10.40 & 141.11 & -18.75 & 10.20 & 4.02 &7.04\\
NGC2778$^{+}$  & SG13 & SB0  & 42.72 & 1.39 & 0.02 & 7.15$^g$ & 13.40 & 10.70 & 154.03 & -18.74 & 10.10 & 3.79& 6.83\\
NGC2787$^{+}$  & N20  & S0   & 7.41  & 0.83 & 0.01 & 7.85$^a$  & -     & -     & 191.92 & -18.45 & 10.14 & 4.30& 6.28\\
NGC3115$^{+}$  & SG13 & S0   & 10.28 & 0.92 & 0.01 & 8.95$^e$  & -     & 11.00 & 260.23 & -20.05 & 10.90 & 4.19 &7.18\\
NGC3115B$^{+}$ & P20  & S0   & 11.00 & 0.56 & 0.07 & -    & -     & -     & 26.0      & -16.97      & 8.95  & 3.12 &6.86\\
NGC3379$^{*}$  & N20  & E    & 6.80  & 0.77 & 0.01 & 8.08$^h$ & 13.40 & 10.61 & 202.47 & -20.16 & 10.83 &3.94& 4.15\\
NGC3384$^{+}$  & SG13 & SB0  & 11.26 & 0.52 & 0.01 & 7.03$^i$ & 7.70  & 10.50 & 144.11 & -19.11 & 10.60 &4.13& 7.34\\
NGC3585$^{+}$  & SG13 & S0   & 17.29 & 0.56 & 0.01 & 8.52$^c$ & -     & 11.20 & 214.20 & -20.97 & 11.10 & 4.10& 6.60\\
NGC3593$^{+}$  & P20  & S0   & 5.50  & 0.68 & 0.01 & 6.38$^j$ & -     & -     & 73.91  & -18.36 & 9.78 & 4.42 &8.20\\
NGC3998$^{+}$  & N20  & S0   & 14.10 & 1.25 & 0.01 & 8.38$^k$ & 11.30 & 10.64 & 265.09 & -19.40 & 10.66 & 4.00 &5.92\\
NGC4026$^{+}$  & SG13 & S0   & 17.83 & 0.74 & 0.01 & 8.26$^e$ & -     & 10.60 & 173.49 & -19.11 & 10.40 & 4.10& 7.11\\
NGC4261$^{+}$  & N20  & E    & 24.00 & 0.81 & 0.01 & 8.70$^a$ & 16.20 & 11.42 & 296.73 & -21.44 & 11.56 & 3.90 &6.32\\
NGC4291$^{+}$  & N20  & E    & 24.30 & 1.26 & 0.01 & 8.51$^a$ & -     & 11.35 & 292.66 & -20.35 & 10.80 & 3.93& 6.70\\
NGC4352$^{*}$  & SG13 & S0   & 19.00 & 0.97 & 0.03 & 6.09$^b$ & 4.50  & 10.30 & 67.58  & -18.04 & 9.80  & 3.52 &6.84\\
NGC4374$^{*}$  & N20  & E    & 10.10 & 0.70 & 0.01 & 9.18$^a$ & 14.90 & 11.28 & 277.59 & -21.73 & 11.56 & 3.76 &7.80\\
NGC4379$^{*}$  & SG13 & S0   & 16.27 & 0.93 & 0.02 & 7.39$^b$ & 2.60  & 10.40 & 110.46 & -18.55 & 10.10 & 3.74 &7.66\\
NGC4387$^{*}$  & SG13 & E    & 17.04 & 0.80 & 0.02 & 6.89$^b$ & 2.70  & 10.20 & 99.95  & -18.47 & 10.10 & 3.75& 7.54\\
NGC4452$^{*}$  & SG13 & S0   & 16.04 & 0.70 & 0.03 & 6.89$^b$ & 4.20  & 9.80  & 99.89  & -      & 10.00 & 3.66 & 6.38\\
NGC4458$^{*}$  & SG13 & E    & 18.39 & 1.12 & 0.02 & 6.83$^b$ & 3.80  & 10.40 & 97.35  & -18.27 & 9.90  & 3.57& -\\
NGC4460$^{*}$  & P20  & S0   & 9.20  & 0.58 & 0.03 & -    & -     & -     & 39.67  & -17.57 & 9.44  & 3.33 & -\\
NGC4476$^{*}$  & SG13 & S0   & 17.69 & 0.78 & 0.02 & 5.93$^b$ & 2.90  & 10.00 & 62.81  & -18.25 & 9.90  & 3.61& 7.05\\
NGC4479$^{*}$  & SG13 & S0   & 18.33 & 0.70 & 0.03 & 6.43$^b$ & 4.30  & 10.00 & 80.48  & -17.83 & 9.80  & 3.70& 6.70\\
NGC4482$^{*}$  & SG13 & dE   & 18.91 & 0.68 & 0.05 & 4.99$^b$ & 4.30  & 10.70 & 41.83  & -17.73 & 9.50  & 3.06& 6.95\\
NGC4486$^{*}$  & N20  & E    & 9.20  & 0.53 & 0.01 & 9.80$^a$ & 17.70 & 11.43 & 323.00 & -21.59 & 11.78 & 3.83& 8.30\\
NGC4550$^{*}$  & SG13 & E/S0 & 16.10 & 0.73 & 0.02 & 6.80$^b$ & 2.20  & 10.20 & 96.25  & -18.65 & 10.10 & 3.72& -\\
NGC4551$^{*}$  & SG13 & E    & 17.01 & 0.84 & 0.02 & 6.93$^b$ & 3.70  & 10.30 & 102.20 & -18.37 & 10.10 & 3.96& -\\
NGC4552$^{*}$  & N20  & E    & 17.10 & 0.86 & 0.01 & 8.70$^a$ & 12.60 & 10.90 & 250.31 & -20.48 & 11.42 & 3.93 &6.96\\
NGC4578$^{*}$  & SG13 & S0   & 17.03 & 1.01 & 0.02 & 7.28$^b$ & 0.53  & 10.80 & 111.92 & -18.90 & 10.30 & 3.84 &7.59\\
NGC4612$^{*}$  & SG13 & S0   & 17.53 & 1.19 & 0.02 & 6.58$^b$ & 2.30  & 10.40 & 85.76  & -19.11 & 10.30 & 3.51 &7.18\\
NGC4623$^{*}$  & SG13 & E    & 16.83 & 0.92 & 0.03 & 6.35$^b$ & 3.90  & 10.00 & 76.99  & -18.15 & 9.90  & 3.66 &-\\
NGC4649$^{*}$  & N20  & E    & 10.20 & 0.54 & 0.01 & 9.32$^a$ & 17.70 & 11.42 & 330.50 & -21.50 & 11.69 & 4.04 &6.30\\
NGC4697$^{+}$  & SG13 & E    & 12.17 & 0.88 & 0.01 & 8.13$^e$ & 11.30 & 11.10 & 165.22 & -20.33 & 11.10 & 3.90 &7.45\\
\hline
\end{tabular}
\end{table*}

\begin{table*}
	\centering
	\centering
	\text{\textbf{Table \ref{tab1}}}\\
	\text{(continued)}
	\begin{tabular}{lccccccccccccr}
	\hline
Galaxy   & Ref. & Type & Dist. & CIR$_I$ & $\bigtriangleup_{CIR_I}$  & M$_{BH}$  & Age & M$_{dyn.}$ & $\sigma$  & M$_B$     & M$_{str}$ & $B-K$ & M$_{NSC}$ \\
         &      &    & (Mpc)      &         &      & (log M$\odot$)     & (Gyr) &          (log M$\odot$) &  (kms$^{-1}$)      & (mag)       & (log M$\odot$)   & (mag) & (log M$\odot$)   \\
\hline
NGC5102$^{+}$  & N20  & S0   & 3.05  & 1.08 & 0.01 & 5.96$^a$ & -     & -     & 61.10  & -17.91 & 9.84  & 3.29& 7.87\\
NGC5389$^{+}$  & SG13 & S0   & 32.20 & 1.04 & 0.02 & -    & -     & 10.60 & 116.13 & -19.77 & 10.60 & 4.25 &-\\
NGC5422$^{*}$  & SG13 & S0   & 28.05 & 1.16 & 0.02 & -    & 9.70  & 10.65 & 160.29 & -19.75 & 10.50 & 4.02 & 7.13\\
NGC5587$^{+}$  & SG13 & S0   & 35.60 & 1.00 & 0.03 & -    & -     & -     & 89.12  & -19.30 & 10.30 & 4.09 &-\\
NGC5689$^{+}$  & SG13 & S0   & 43.30 & 1.06 & 0.01 & -    & -     & -     & 150.02 & -20.51 & 10.80 & 4.31 &-\\
NGC5838$^{+}$  & SG13 & E/S0 & 19.72 & 1.02 & 0.01 & 8.00$^l$ & 11.30 & 11.20 & 273.56 & -20.04 & 10.70 &4.17&7.67 \\
NGC5854$^{+}$  & SG13 & S0   & 14.62 & 0.93 & 0.02 & -    & 4.30  & 10.30 & 102.36 & -19.81 & 10.40 & 3.81 &7.98\\
NGC6010$^{+}$  & SG13 & S0   & 21.10 & 1.24 & 0.02 & -    & 12.00 & 10.52 & 147.84 & -20.00 & 10.50 & 4.21& 7.53\\
NGC7457$^{+}$  & SG13 & S0   & 36.03 & 0.87 & 0.02 & 6.95$^m$ & -     & 9.89  & 67.97  & -19.14 & 10.20 & 3.65&7.50\\
UGC7399A$^{+}$ & SG13 & dE   & 16.90 & 0.53 & 0.07 & 5.33$^b$ & 7.10  & 9.80  & 43.06  & -17.22 & 9.10  & 2.50&6.59\\
UGC7436$^{*}$  & SG13 & dE   & 16.04 & 0.56 & 0.07 & 4.40$^b$ & 5.90  & 9.81  & 32.17  & -16.60 & 9.10 & 2.81&5.81\\
\hline	
\end{tabular}
\end{table*}

\section{Variation of CIR$_I$ with host galaxy properties}
\label{sec:results}
CIR is found to be an important parameter in galaxy evolution studies manifested through different correlations and trends reported in the literature \citep[e.g.][]{2018MNRAS.477.2399A,2020RAA....20...15A,2021MNRAS.500.1343S}. These studies suggest the use of CIR as a tool to explore the co-evolution between various structural and dynamical properties of different Hubble types of galaxies. This paper discusses the variation of CIR$_{I}$ with different host galaxy properties for a sample of 63 nearby ETGs hosting NSCs.  

\begin{figure*}
	\includegraphics[width=1.0\textwidth]{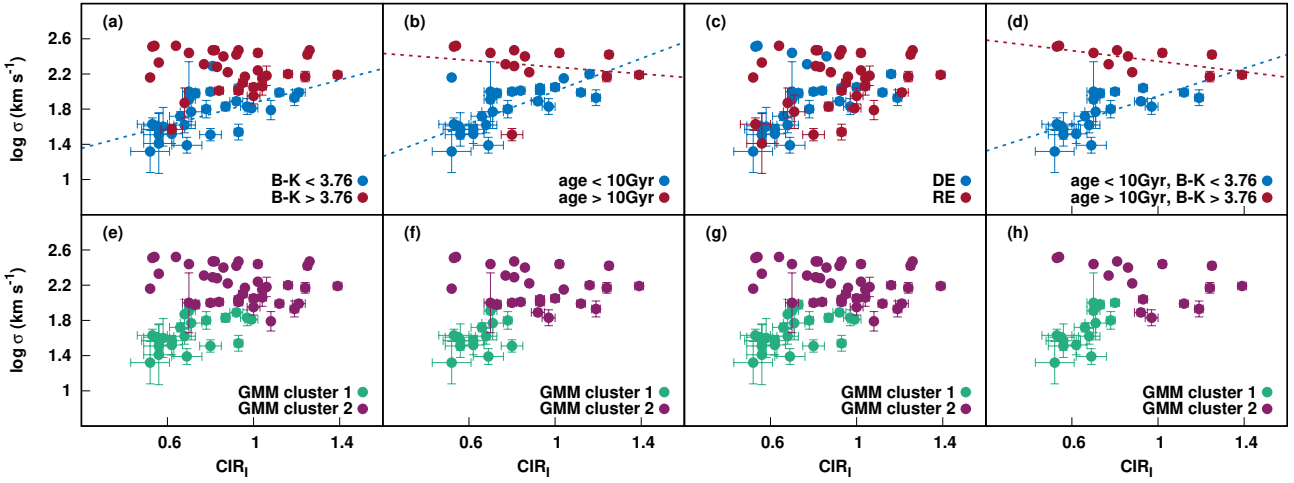}
   \caption{Variation of CIR$_I$ with central velocity dispersion, $\sigma$, with colouring schemes based on $B-K$ (panel {\bf a}), age (panel {\bf b}), environment (panel  {\bf c}) and $B-K$ and age together (panel {\bf d}) of the sample galaxies. In panel {\bf a}, blue galaxies ($B-K$ $<$ 3.76) are denoted with blue filled circles whereas red galaxies  ($B-K$ $>$ 3.76) are given in red filled circles. In panel {\bf b}, young galaxies (with age less than 10 Gyr)  are shown in blue filled circles while old galaxies (age greater than 10 Gyr) are marked with red filled circles. In panel {\bf c}, galaxies from the dense environment (DE) are denoted with blue filled circles and galaxies from the rare environment (RE) are represented as red filled circles, respectively. Please see the text for the criteria used for defining the environment. In panel {\bf d}, blue ($B-K$ $<$ 3.76) as well as young (with age less than 10 Gyr) galaxies are represented in blue filled circles whereas red ($B-K$ $>$ 3.76) and old (age greater than 10 Gyr) galaxies are denoted with red filled circles. The bottom panel shows the distribution of two clusters statistically identified using GMM.  GMM cluster 1 galaxies are denoted with green filled circles whereas GMM cluster 2 galaxies are given in purple filled circles. In the upper panel, the best fit linear relations for the sub-samples separately are shown as dotted lines where correlations are significant.}
    \label{multi-sigma}
\end{figure*}

\subsection{Variation of CIR$_I$ with central velocity dispersion}
In Fig.\ref{multi-sigma}, variation of CIR$_I$ with central  velocity dispersion is shown. Central stellar velocity dispersion is a tracer of galaxy mass validated by various studies \citep[e.g.][]{2010ApJ...724..511A}. It is well known that, the luminosity as well as Mg$_{2}$ index of a galaxy tightly correlates with velocity dispersion \citep[e.g.][]{1976ApJ...204..668F, 1999MNRAS.303..813C}. Likewise, a tight correlation exists between the masses of supermassive black holes and the velocity dispersions of the bulges that host them \citep{2000ApJ...543L...5G}. It is also reported to exhibit a significantly strong correlation with the total
gravitating mass of a galaxy making it a very fundamental parameter in galaxy evolution studies \citep{2015ApJ...800..124B}.  Even though no apparent connection between CIR and velocity dispersion is seen, there appears the possibility of at least two populations showing different trends. This is not completely surprising. Owing to the simple definition of CIR, higher will be its value  when the (projected) intensity in the inner aperture is increased (by virtue of, say, increased star formation near the centre) or when light contribution in the outer shell ($r_2$ - $r_1$) is decreased (as can happen when obscured by dust).   Hence it's quite natural to expect the CIR to be less effective when a sample contains sets with different characteristics that can affect the value of CIR differently.  In order to explore this, in Fig. \ref{multi-sigma}({\bf a}), we split the sample into blue and red ones using the median $B-K$ colour (3.76) of the sample galaxies. The median value was used for minimising  uncertainties associated with the low number statistics in a group.  $B-K$ colour is a good indicator of the amount of dust extinction \citep{1995A&A...300L...1P}. Galaxies bluer than the median $B-K$ colour (3.76), hereafter blue galaxies, are represented as blue filled circles and those redder than the median,  hereafter red galaxies,  are shown as red filled circles. Blue galaxies show a mild positive trend (linear correlation coefficient, $r = 0.64$ with a significance, $S = 99.95$ per cent, see Table \ref{tab2}) in the plot suggesting an increase in galaxy mass as CIR$_{I}$ increases while red galaxies in our sample do not show any apparent association. The nature of clustering in the sample is independently verified using  Gaussian Mixture Model analysis \citep[GMM;][]{2011ApJ...729..141B,2022A&A...660A..22W}.  A basic assumption of GMM is that all the data points are drawn from a mixture of a finite number of Gaussian components with unknown distribution parameters \citep{2018AJ....156..121G}. The GMM can assign a membership score (posterior probability) to each data point that describes how similar each point is to each cluster's centre; which is an automated process and does not require manual intervention. GMM is being used in various fields, including astrophysics \citep{2017MNRAS.469.3374C,2018ApJ...855...14K}, as an unsupervised clustering technique. The representation of regularly distributed subpopulations within a population can be done probabilistically using GMM. This statistical model uses an Expectation Maximization (EM) algorithm \citep{10.2307/2984875} to optimize the model's likelihood. In Fig. \ref{multi-sigma}({\bf e}) the different groups identified by  GMM are given in green and purple filled circles. The result of GMM also suggests the presence of two groups within the sample which are statistically different.

In Fig. \ref{multi-sigma}({\bf b}), we use age as a parameter  to explore the possible  subsets. Here SSP age reported in literature \citep{2009MNRAS.394.1229C,2014ApJ...783..135A,2016MNRAS.460.4492D} is used.  Galaxies with an age less than 10 Gyr (hereafter young galaxies) are denoted as blue filled circles and those with an age more than 10 Gyr (hereafter old galaxies)  are represented as red filled circles.  Here young galaxies follow a positive trend ($r = 0.79$,  $S > 99.99$ per cent), while the old ones appear to possess a negative trend ($r = -0.77$,  $S = 99.45$ per cent) though not reliable due to the limited number of galaxies. Here also the GMM analysis identifies clusters similar to the ones we constructed using age (Fig. \ref{multi-sigma}({\bf f})).
 
Next, we explore the influence of environment of galaxies. In order to do that a galaxy is considered in the rare environment if it possesses a surrounding galaxy density  below 36 Mpc$^{-3}$ within 0.5 Mpc in the plane of the sky and if no nearest neighbour falls within 250 km\,s$^{-1}$ in recession velocity, otherwise it is treated from dense environment\footnote{Determined using the link \url{https://ned.ipac.caltech.edu/forms/denv.html}.}.  In  Fig. \ref{multi-sigma}({\bf c}), galaxies from the dense environment (hereafter DE) are given as blue filled circles whereas those from the rare environment (hereafter RE) are shown as red filled circles.    No apparent trends are visible here. However, the statistical clusters identified by GMM, shown in Fig. \ref{multi-sigma}({\bf g}), are not comparable here suggesting the environment of a galaxy does not appear to be a good differentiator in this sample.

Application of both  age and colour criteria together, however,  reveals the presence of two distinct sets of galaxies present in the sample and is shown in Fig. \ref{multi-sigma}({\bf d}).  It is evident from the Figure that, young and dusty systems seem to follow a positive trend ($r = 0.73$, $S = 99.94$ per cent).  On the other hand, old and less dusty systems tend to follow a near-perfect negative correlation ($r = -0.86$,  $S = 99.73$ per cent). As shown in Fig. \ref{multi-sigma}({\bf h}), the application of GMM also reveals two statistical sets of galaxies. The central velocity dispersion of galaxies enjoys a very strong and positive correlation with the mass of central SMBH.  The mass of SMBH of most of our sample galaxies is estimated using their $\sigma$ values. Hence all the  correlations/trends with $\sigma$ mentioned here can stand as a proxy to those involving the mass of SMBH as well.

\begin{table}
\caption{The Table lists the linear correlation coefficients ({\rm r}) along with significance ({\rm S}) for various relations given in Fig.\ref{multi-sigma} and Fig.\ref{multi-cir}. N denotes the number of galaxies. For completeness, the table also includes relations that show little correlation. The sub-samples are defined in section \ref{sec:results}.}
\label{tab2}
\begin{adjustbox}{width=\columnwidth}
\begin{tabular}{lllll}
Correlation                      & Sub-sample    & r     & S                  & N  \\
\hline
\multirow{8}{*}{CIR - log $\sigma$} & Blue          & \ 0.64  & 99.95              & 28 \\
                                 & Red           & -0.30  & 87.78              & 30 \\
                                 & Young         & \ 0.79  & \textgreater 99.99 & 26 \\
                                 & Old           & -0.77 & 99.45              & 13 \\
                                 & DE            & \ 0.20   & 67.68              & 28 \\
                                 & RE            & \ 0.23   & 79.53              & 33 \\
                                 & Blue \& Young & \ 0.73  & 99.94              & 20 \\
                                 & Red \& Old    & -0.86 & 99.73              & 11 \\
\hline
\multirow{2}{*}{CIR - M$_B$}         & Blue          & -0.65 & 99.98              & 25 \\
                                 & Red           & \ 0.68   & 99.97              & 27 \\
\hline                                 
\multirow{2}{*}{CIR - M$_{dyn}$}       & Blue          & \ 0.63  & 99.91              & 26 \\
                                 & Red           & -0.33 & 88.42              & 26 \\
\hline                                 
\multirow{2}{*}{CIR - M$_{str}$}        & Blue          & \ 0.64  & 99.98              & 30 \\
                                 & Red           & -0.30  & 89.1               & 32 \\
\hline                           
\multirow{2}{*}{CIR - M$_{nsc}$}        & Blue          & \ 0.71  & 99.99              & 27 \\
                                 & Red           & \ 0.01  & 3                  & 28 \\
\hline                                 
\end{tabular}
\end{adjustbox}
\end{table}
\subsection{Variation of CIR$_I$ with M$_{B}$}  

\begin{figure*}
	\includegraphics[width=0.8\textwidth]{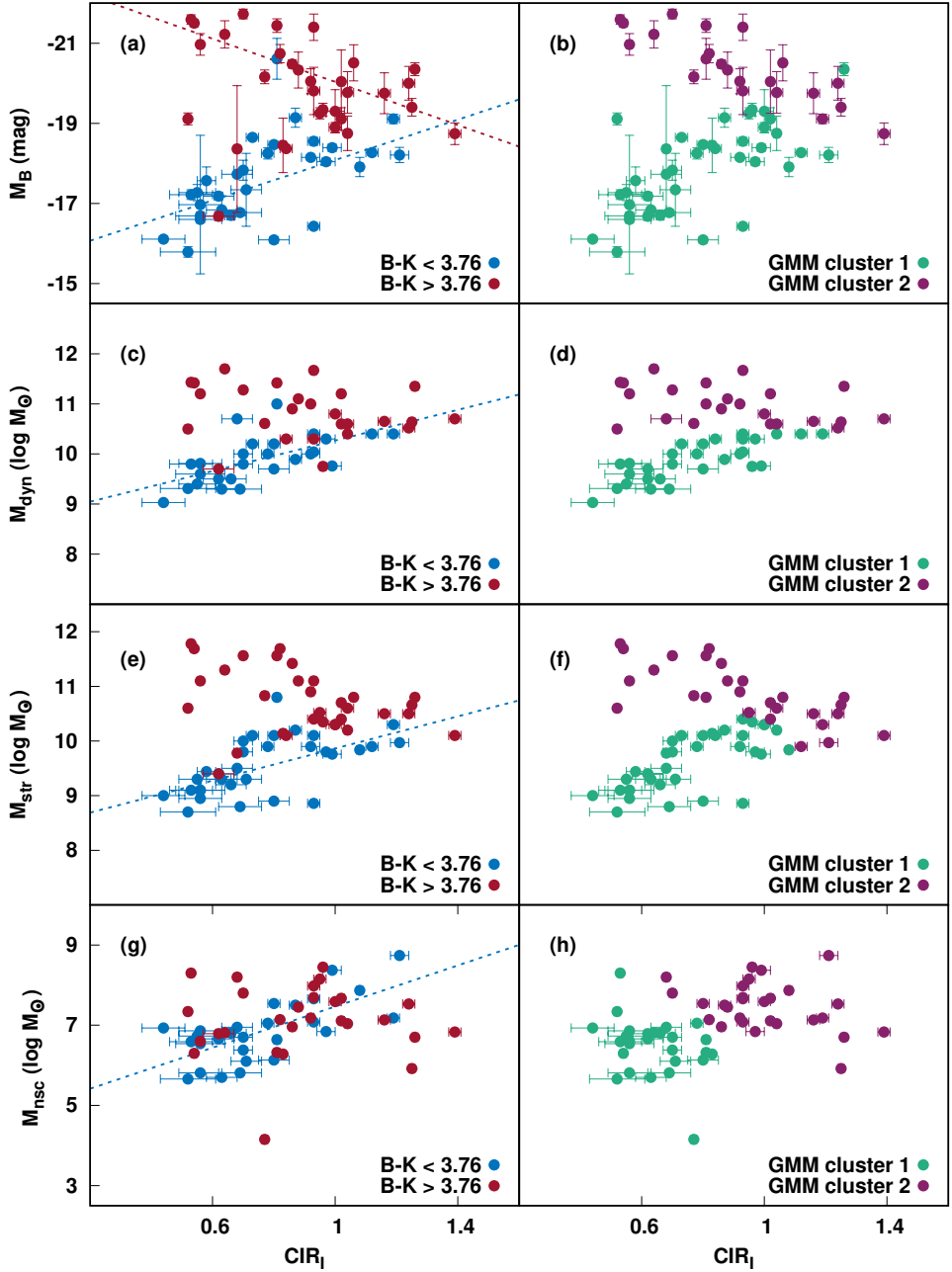}
   \caption{Left panel shows the variation of CIR$_I$ with absolute B band magnitude of the host galaxy (panel {\bf a}),  dynamical mass of the galaxy M$_{dyn}$ (panel {\bf c}), stellar mass of the host galaxy, M$_{str}$(panel {\bf e}) and  mass of NSC of the host galaxies, M$_{nsc}$ (panel {\bf g}). Galaxies bluer than the median of $B-K$ colour are denoted with blue filled circles whereas galaxies redder than the median of $B-K$ colour are represented as red filled circles. The right panel shows the distribution of two clusters identified statistically using GMM.  Galaxies in GMM cluster 1 are marked with green filled circles, whereas galaxies in GMM cluster 2 are marked with purple filled circles. In the left panel, the best fit linear relations for the sub-samples separately are shown as dotted lines where correlations are significant.}
    \label{multi-cir}
\end{figure*}

Variation of absolute B band magnitude with CIR$_{I}$ is shown in Fig. \ref{multi-cir}({\bf a}). Galaxies bluer than the median  $B-K$ colour are denoted with blue filled circles whereas the redder galaxies are represented using red filled circles. Blue  galaxies in the sample show a mild positive trend ($r = -0.65$, $S = 99.98$ per cent) whereas red  galaxies show a weak negative trend ($r = 0.68$, $S = 99.97$ per cent). Optical B band is particularly sensitive to star-forming components in low redshift galaxies \citep{2003MNRAS.346..304J}.  Most of the blue galaxies are faint (with M$_{B}< -19 $), while red galaxies are aligned with bright galaxies in the sample.  As the blue (and faint) galaxies may contain more dust (in the central region) than their red counterparts, an increased infra-red emission from their centres is possible \citep{2010MNRAS.403.1894D} which can rise their CIR$_{I}$ value.  On the other hand, a black hole's ability to remove dust from its sphere of influence \citep{2017Natur.549..488R} and associated quenching may be the primary physical driver of the negative trend shown by red galaxies. A very similar clustering is obtained using GMM and is shown in Fig. \ref{multi-cir}({\bf b}) where the green filled circles represent the galaxies in GMM cluster 1, while the purple filled circles represent the galaxies in GMM cluster 2.

\subsection{Variation of CIR$_I$ with M$_{dyn}$}

The dynamical mass (M$_{dyn}$) is a galaxy parameter that depends on velocity dispersion and size of the galaxy \citep{2010ApJ...724..511A}. For the sample galaxies,  M$_{dyn}$ estimations are adopted from \citet{2013ApJ...763...76S} or \citet{2016MNRAS.460.4492D}.   However, they have used different methods to estimate it.  \citet{2013ApJ...763...76S} used virial estimator M$_{dyn}$ = $\alpha\sigma_{e}^{2}$R$_e$/G, where R$_e$ is the effective half-light radius and $\sigma_{e}$, is the luminosity-weighted velocity dispersion measured within a \mbox{1R$_e$} aperture. \citet{2016MNRAS.460.4492D} calculated M$_{dyn}$ using the following equation;
\begin{equation}
    M_{dyn} = \frac{K_V}{G} R_e \sigma_0^2,
\end{equation} 
where \emph{K$_V$} is a factor that depends on the shape of the density profile of the ETG, \emph{G} is the gravitational constant, \emph{R$_e$} is the effective radius of the galaxy and \emph{$\sigma_0$} is the velocity dispersion at the centre of the galaxy. The variation of M$_{dyn}$ is plotted against CIR$_{I}$ in Fig. \ref{multi-cir}({\bf c}) and clusters identified using GMM are shown in Fig. \ref{multi-cir}({\bf d}). Albeit the heterogeneity in the method of determination of M$_{dyn}$, here also we can observe that the blue galaxies show a weak positive trend ($r = 0.63$, $S = 99.91$ per cent) with CIR$_{I}$. For the red systems in the plot, no trend is visible.  Dust content in galaxies is reported to possess good correlations with stellar mass \citep{2014A&A...562A..30S} and star formation rate \citep{2010MNRAS.403.1894D,2014A&A...562A..30S}. In a galaxy with a high dust content, the star formation rate can increase, causing the CIR$_{I}$ to increase with its dynamical mass. 
 
\subsection{Variation of CIR$_I$ with M$_{str}$}
Stellar mass  of the sample galaxies, M$_{str}$, were compiled from \citet{2013ApJ...763...76S}, \citet{2020A&ARv..28....4N}, \citet{2020ApJ...900...32P} and \citet{2021A&A...650A.137F}. \citet{2020ApJ...900...32P} used \emph{Spitzer} 3.6$\mu$m luminosities and a mass-to-light ratio (M/L) of 0.5 at 3.6$\mu$m to estimate the galaxy stellar masses. \citet{2013ApJ...763...76S} obtained stellar mass from the total galaxy luminosity in $K$ band, M$_{K}$ (using 2MASS observations), assuming a standard mass-to-light ratio, M/L$_{K}$ = 0.8 \citep[e.g.][]{2007ApJ...665.1084B,2001ApJ...550..212B}.  Fig. \ref{multi-cir}({\bf e}) shows  the variation of M$_{str}$  with CIR$_{I}$ and Fig. \ref{multi-cir}({\bf f}) shows the same using GMM clustering. Here also the blue galaxies show a systematic increase in CIR$_{I}$ with increasing stellar mass ($r = 0.64$, $S = 99.98$ per cent) which is not completely surprising as dusty galaxies can harbour high  star formation  \citep[e.g.][]{2012A&A...540A..52C,2012MNRAS.421.3027B}. For the red systems in the plot, however, no apparent association can be found.
 
\subsection{Variation of CIR$_I$ with M$_{nsc}$}

The variation of CIR$_I$ with M$_{nsc}$ is shown in Fig. \ref{multi-cir}({\bf g}).  Blue galaxies show a slight positive trend ($r = 0.71$, $S = 99.99$ per cent) whereas red galaxies seem to be scattered. The statistical cluster association obtained using GMM, shown in Fig. \ref{multi-cir}({\bf h}), does not show any similarity with that obtained using $B-K$ colour. The masses of NSCs obey correlations with properties of the host galaxies such as velocity dispersion and bulge mass \citep{2006ApJ...644L..21F}.  Early-type galaxies exhibit a low contrast between NSC and host-galaxy body making the spectroscopic studies of NSCs in ETGs extremely difficult \citep{2008JPhCS.131a2043B}. Most of the existing mass estimates of NSCs are photometric,  based generally on SED fitting or colour-mass-to-light ratios of galaxies. These may not be the best indicators of mass due to age-metallicity degeneracy and because NSCs contain complex star formation history \citep{2020ApJ...900...32P}.  Despite these,  CIR$_{I}$  seems to vary  with M$_{nsc}$ for dusty systems in the sample.  The CIR$_{I}$, estimated using  \emph{Spitzer} space telescope, appears to trace many properties of  galaxies for dusty systems in the sample.

\section{DISCUSSION AND CONCLUSION}
\label{sec:discussion}
In this paper, we determine the central intensity ratio of early-type galaxies hosting nuclear star clusters using near-infrared light and present its variations with different host galaxy properties such as central velocity dispersion ($\sigma$), absolute B band magnitude (M$_{B}$), dynamical mass (M$_{dyn}$), stellar mass (M$_{str}$) and mass of nuclear star cluster (M$_{nsc}$) of the host galaxy and reflect the possible connection between the evolution of galaxies and their central light distribution.

The values of CIR measured from nearby bands are known not to disturb correlations or trends, involving CIR \citep{2021MNRAS.500.1343S}. However, CIR measured from different broadbands can vary differently. In the optical band, the dust present in the (projected) central region of galaxies can limit the light visible to observers. On the other hand, in near-infrared, one gets dust-penetrated light from the centres of galaxies which may originate from small dust grains that are heated by energetic photons produced by young stars \citep{2013A&A...558A.136M}.  Dusty galaxies in the sample where CIR$_{I}$ shows a positive trend with M$_{nsc}$, thus suggesting a mutual growth of NSCs with CIR$_{I}$. It is likely that NSCs are  part of star formation near the centre in low-mass galaxies. The growth of NSCs in the centers of these galaxies may add light to their centers. However,  high-mass galaxies in the sample do not show such a trend.
Variation of M$_{nsc}$ with the host galaxy parameters such as central velocity dispersion, absolute B band magnitude, dynamical mass, and stellar mass of the sample galaxies seem to be scattered.  Moreover, CIR$_{I}$ shows a positive trend with all the above-mentioned parameters for low mass, blue, young, and faint galaxies in the sample. However, in high-mass galaxies in the sample the increased mass of SMBH might have reduced the NSC domination in the centre, resulting in the lack of any correlation with CIR$_I$.

CIR$_{I}$ is showing a positive trend with $\sigma$ for young and dusty galaxies in the sample suggesting an increase in dust and light content corresponding to the emission of 3.6$\mu$m band as the galaxy grows, while their older counterparts show no significant trend. Environment  seems insignificant in the CIR$_{I}$ - $\sigma$ plot though $\sigma$ is coupled with the local galaxy density \citep{2017MNRAS.468..333S}. Dust abundance is directly connected with galaxy growth through the formation of new stars \citep{2014A&A...562A..30S}. Thus, it is possible to attribute the growth in central velocity dispersion to  that in galaxy mass.

An increase in $\sigma$ also indicates the possibility of growth in the size of the galaxy's central black hole, its central potential, and its total mass validated by various studies \citep[e.g.][]{2010ApJ...724..511A,2000ApJ...543L...5G}. \citet{2018MNRAS.477.2399A}, hereafter AR18, reported that CIR calculated in the optical band (CIR$_{V}$) shows a strong negative correlation with the mass of the supermassive black hole and central velocity dispersion for classical bulges, whereas that of spirals and lenticulars with pseudo-bulges remained outliers. 8 ellipticals and 6 lenticulars in AR18 are present in our sample also.  The pseudo-bulges,  which behaved as outliers in AR18,  seem to obey a positive trend here along with the blue galaxies whereas classical bulges in their correlation seem to be a part of the red galaxies in our sample.  \citet{2021MNRAS.500.1343S} found that, CIR$_{V}$ follows negative correlations with M$_{BH}$ and $\sigma$ for field ETGs also. These studies discussed the possible quenching that happened in these systems as the black hole grows. In our sample, CIR$_{V}$ does not show any significant correlation with M$_{BH}$ or $\sigma$. From Fig.\ref{multi-sigma}({\bf a}), it is evident that the majority of our sample galaxies are carrying dust in them. The presence of dust can disturb the relation in the optical band. The correlations/trends described by CIR are context-dependent. In the studies of \citet{2018MNRAS.477.2399A} and \citet{2021MNRAS.500.1343S}, the correlations involving CIR$_V$ were possible as most of their sample galaxies contained low amounts of dust. In the current study, on the contrary, most of the NSC host galaxies are reported to contain high amounts of dust, making observations in the near-infra-red band  well suited to study their central light distribution. Interstellar dust in galaxies absorbs energy from starlight; this absorbed energy is then re-radiated at infra-red and far-infra-red wavelengths \citep{2007ApJ...657..810D}. The contribution of thermal dust to the observed emission at 3.6$\mu m$ can be powered by AGN and/or star formation \citep{2011MNRAS.413.1479S}. \citet{2009MNRAS.397.2148G} discuss two possible ways of BH growth in galaxies hosting NSCs, (i) some massive BHs may grow through runaway collision of NSC stars, (ii) inward gas flow may result in star formation at the galactic centre, that itself may be fuelling the growth of BHs. It is likely that the same gas that is feeding star formation also feeds the black hole \citep{2010MNRAS.407.1529H}. In that sense, the presence of an NSC may, in fact, enhance the growth rate of the central black hole \citep{2020A&ARv..28....4N}. Nuclear star formation is more solidly coupled to the growth of central black holes than the global star formation rate \citep{2010MNRAS.407.1529H}. Dusty galaxies in our sample reveal that their central light distribution is coupled to NSC growth. If NSC growth can boost central BH growth, it is natural to assume that there may be a growth in the central potential also. Thus our analysis supports the co-evolution of NSCs with their host galaxies reported in the literature, at least for dusty galaxies.

\begin{figure}
	\includegraphics[width=1.0\columnwidth]{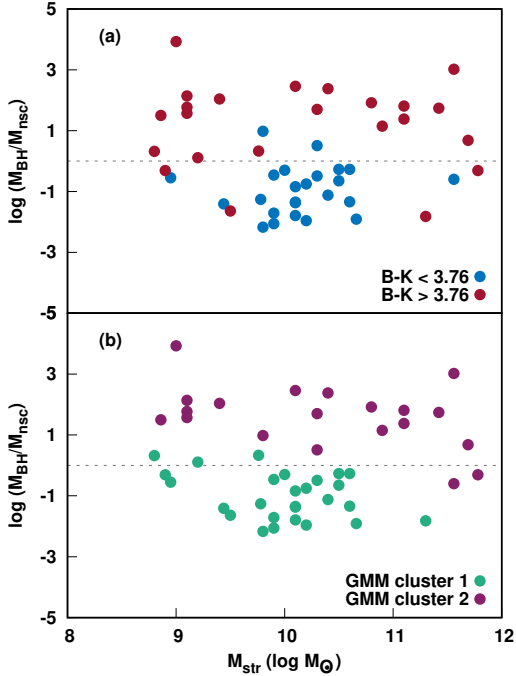}
   \caption{The variation of the ratio of the masses of the central massive black hole and the NSC with their host galaxy stellar mass. The top panel (panel \textbf{a}) gives $B-K$ clustering and the bottom panel (panel \textbf{b}) gives GMM clustering results of the sample galaxies. The dashed line indicates equal-mass black holes and NSCs in galaxies, while objects above the line have more massive black holes than NSCs. Symbols of galaxies are the same as shown in Fig.\ref{multi-sigma}.}
    \label{fraction_stellarmass}
\end{figure}

\citet{2006ApJ...644L..21F} reported that, in low-mass galaxies (Galaxy mass < 10$^{10}$ M$_\odot$), which may not have supermassive black holes (searches are ongoing for low-mass central BHs ($\leq$ 10$^{6}$ M$_\odot$), the masses of nuclear star clusters correlated with the one-dimensional velocity dispersion ($\sigma_{bulge}$). However, the connection  between NSCs and central BHs is not well understood. The sample of galaxies with both a detected NSC and a BH is quite limited. Even from this data, it is clear that, low-mass (10$^{9}$ - 10$^{10}$ M$_\odot$) galactic nuclei are dominated by NSCs while high-mass (Galaxy mass > 10$^{10}$ M$_\odot$) galaxy nuclei are dominated by SMBHs \citep{2018ApJ...858..118N}. Increasing evidence also points out that, massive star clusters are sites of inter-mediate mass black hole formation \citep{2006AJ....132.2539S}. But direct observational pieces of evidence for such BHs are less since such small BHs are extremely difficult targets for dynamical searches \citep{2012AdAst2012E..15N}.  Nuclear star clusters are common in those spheroids with stellar mass M$_{str}$ $\sim$ 10$^{8-11}$ M$\odot$ harbouring a M$_{BH}$ < 10$^{7}$ M$\odot$ and may contribute significantly to the mass of central blackhole. A larger value of M$_{BH}$ can prevent the co-existence with NSC because the very central stars should be either disrupted by the BH gravity or kicked away by tidal forces \citep{2011MNRAS.413.1479S}. These interactions with NSCs and BHs possibly explain the negligible nucleation fraction for galaxies with M$_{str}$ > 10$^{11}$ M$\odot$ \citep{2021A&A...650A.137F}. The reason for massive galaxies showing deviation from the positive trend exhibited in CIR$_{I}$ - $\sigma$ plot  can be explained using the above-proposed interactions. \citet{2012AdAst2012E..15N} discuss two possible scenarios that can happen in high mass galaxies, \mbox{(i) either} the galaxies with massive BHs never had a sizeable NSC, possibly because their central BHs grew early in the age of the universe; thus stopping NSC growth, or, (ii) massive BHs destroy their host NSCs. The positive CIR$_{I}$-$\sigma$ trend in Fig. \ref{multi-sigma} possibly suggests the simultaneous growth of NSC and BH. However, we can not make a strong statement on red galaxies in the sample as no such trends are visible in  CIR$_{I}$-$\sigma$ variation. 

A positive correlation is reported between M$_{str}$ and $\log$ (M$_{BH}$/M$_{nsc}$) for a sample of galaxies hosting NSC \citep{2020A&ARv..28....4N}.  They applied very strict conditions to construct their limited sample,  such as dynamically estimated masses for BHs and NSCs.  However, no such correlation is observed for this study, as can be seen in  Fig. \ref{fraction_stellarmass}, where we used all early-type galaxies hosting NSCs.  In the figure, a comparison of $B-K$ clustering and GMM clustering is given. In panel \textbf{a}, blue galaxies are shown in blue filled circles whereas red galaxies are denoted as red filled circles. In panel \textbf{b}, galaxies in GMM cluster 1 are denoted by green filled circles, and galaxies in GMM cluster 2 are represented by purple filled circles. The dashed line represents the equal mass NSCs and BHs in galaxies. Galaxies above the dashed line host more massive BHs than NSCs while galaxies below the dashed line host BHs less massive than NSCs. It can be seen that in the red galaxies in the sample, the central region is  dominated by the (central) blackhole, where as it is the nuclear star cluster that dominates the centre of blue galaxies. The GMM clustering also reproduces the result.

Using simple photometric techniques, we analyzed centers of nearby ETGs hosting NSCs. In these galaxies, NSCs appear to be involved in central star formation. Improved mass determinations of NSCs in the future may allow us to study galaxies hosting NSCs using CIR$_{I}$. CIR$_{I}$ can be used as a proxy to measure central velocity dispersion in dusty galaxies where AGN/heavy star formation prevents its measurement. In this context, analysis involving the simple photometric quantity CIR seems promising to shed more light on the galaxy evolution scenarios.

\section*{Acknowledgements}
We thank the anonymous reviewer for his/her comments that greatly improved the content of this paper. KS would like to acknowledge the financial support from INSPIRE program conducted by the Department of Science and Technology (DST), Government of India.  KS wishes to thank UGC-SAP and FIST 2 (SR/FIST/PS1159/2010) (DST, Government of India) for the research facilities in the Department of Physics, University of Calicut.  This work is based on archival data obtained with the \emph{Spitzer} Space Telescope, which was operated by the Jet Propulsion Laboratory, California Institute of Technology under a contract with NASA. Support for this work was provided by NASA through an award issued by JPL/Caltech.   We acknowledge the use of the HyperLeda database  (\url{http://leda.univ-lyon1.fr}) and the NED, NASA/IPAC Extragalactic Database (\url{http://ned.ipac.caltech.edu}), which is operated by the Jet Propulsion Laboratory, California Institute of Technology, under contract with the National Aeronautics and Space Administration. This research has made use of NASA's Astrophysics Data System Bibliographic Services. 
\section*{Data Availability}
The near-infra-red observations used in this work are freely accessible on the \textit{Spitzer} Heritage Archive. The data presented in each figure will be shared on reasonable request to the corresponding author.



\bibliographystyle{mnras}
\bibliography{cir_nsc} 








\bsp	
\label{lastpage}
\end{document}